\documentstyle[12pt]{article}
\def\theequation{\arabic{section}.\arabic{equation}} % used in equation labels!
\newcommand{\beq}{\begin{equation}}
\newcommand{\eeq}{\end{equation}}

\begin{document}

\title{ \bf Decoherence, Correlation, and Unstable Quantum States in
Semiclassical Cosmology}
\author{Mario Castagnino$^{1,2,3}$ and Fernando Lombardo$^{1,2}$ \\
\\
$^1$RggR, Service de Chimie-Physique, Universit\'e Libre de Bruxelles\\
Campus Plaine 231, 1050 Bruxelles, Belgium \\
$^2$Departamento de F\'\i sica, Facultad de Ciencias Exactas\\  y Naturales
Universidad de Buenos Aires, Ciudad Universitaria \\ 1428 Buenos Aires,
Argentina \\
$^3$Instituto de Astronom\'\i a y F\'\i sica del Espacio \\ Casilla de Correo
67, Sucursal 28, \\ 1428 Buenos Aires, Argentina}

\thispagestyle{empty}
\vspace{3cm}

\maketitle
\begin{abstract}
\noindent
It is demonstrated that almost any S-matrix of quantum field theory in curved
spaces posses an infinite set of complex poles (or branch cuts). These poles
can
be transformed into complex eigenvalues, the corresponding eigenvectors being
Gamow vectors. All this formalism, which is heuristic in ordinary Hilbert
space,
becomes a rigorous one within the framework of a properly chosen rigged Hilbert
space. Then complex eigenvalues produce damping or growing factors. It is known
that the growth of entropy, decoherence, and the appearance of correlations,
occur in the universe evolution, but only under a restricted set of initial
conditions. It is proved that the damping factors allow to enlarge this set
up to almost any initial conditions.

\end{abstract}

%%%%%%%%%%%%%%%%%%%%%%%%%%%%%%%%%%%%%%%%%%%%%%%%%%%%%%%%%%%%%%%%%%%%%%%%%%%%%%
%%%%%%%%%%%%%%%%%%%%%%%%%%%%%%%%%%%%%%%%%%%%%%%%%%%%%%%%%%%%%%%%%%%%%%%%%%%%%%
%%%%%%%%%%%%%%%%%%%%%%%%%%%%%%%%%%%%%%%%%%%%%%%%%%%%%%%%%%%%%%%%%%%%%%%%%%%%%%
\newpage

\setcounter{page}{1}

\section{Introduction}
For many years unstable quantum states were represented by Gamow vectors
\cite{1}, i.e., eigenvectors corresponding to complex eigenvalues of the
hamiltonian. But since the hamiltonian is self adjoint,
if we use a Hilbert space as state space, eigenvalues must be real.  For this
reason, Gamow vectors were deshonorably excluded from ordinary quantum
mechanics
and they were considered just as useful (but not rigorous) analogies or
approximations. Nevertheless some years ago it was proved that Gamow vectors
belong to an extension of Hilbert space, namely a rigged Hilbert space with a
nuclear subspace based in  Hardy class functions (see e.g.\ \cite{2,3,4} and
bibliography therein). Since then, Gamow vectores are legal citizens of an
extended version of quantum mechanics, where complex eigenvalues greately help
the computation of survival probabilities, life times, Liapunov variables, the
evolution toward equilibrium,  etc. \cite {5,6}. These eigenvalues also
allow the introduction of more refined physical concepts, e. g. the
thermodynamical arrow of time can be defined in a way which is free of the
usual
criticisms (Lochshmidt objection, coarse-graining ambiguities, non-systematic
approximations, etc.). Precisely, studying this arrow of time we can see that
all the problem of time asymmetry essentialy has a cosmological origin
\cite{5,6}. Therefore it is natural to demand if unstable quantum states,
considered as vectors of a rigged Hilbert space, can be used in quantum
cosmology. This is, in fact, the case and the first examples are Refs \cite {7}
and \cite {8} where a simplified (toy model) version of the universe is studied
as a Friedrichs model [4].

In this paper we will present a complete (not only a toy model)
semiclassical model of the universe following the line of Refs.\cite {9} and
\cite {10} and we will show how the presence of unstable quantum states
enlarges the set of cases where we can prove that the decoherence phenomena
appears. Also correlations appear (for unstable states) explaining the outcome
of
a classical universe. Essentially, in this paper,  we study, just
one example, but we will also especulate about eventual generalizations.

But first, let us briefly recall the guiding lines of the extension from
Hilbert space to rigged Hilbert space in usual quantum mechanics. The
traditional
set of states of a quantum system is a  Hilbert
space, which leads,  as it is well known, to time-reversibility. It is
precisely
this property which changes drastically with the extension of the Hilbert space
$\cal{H}$  to a rigged Hilbert space. This extension corresponds essentialy to
the transition from a space of
 square integrable functions to a space of distributions. This procedure is not
unique and different distribution spaces can be defined which are  based on
different test function spaces. If we choose as the test function space
$\Phi_-$,   generated by the eigenfunctions of the energy $\omega$ which are
analytic in the lower complex halfplane, when the real variable
$\omega$ is promoted to a complex variable $z$ (precisely Hardy class
functions), we obtain the dual space $\Phi^\times_-$, which is the required
extension of the Hilbert space $\cal{H}$. The corresponding Gel'fand triplet is
then

\begin{equation}\Phi_- \subset {\cal H} \subset \Phi^{\times}_-.\end{equation}

If the same procedure is performed in the upper complex plane, the resulting
triplet reads

\begin{equation}\Phi_+ \subset {\cal H} \subset
{\Phi^{\times}_+}.\end{equation}

As we will see the first of these choices, hence the space $\Phi^{\times}_-$,
corresponds to unstable dacaying states while the second one, namely
$\Phi^{\times}_+$,  corresponds to unstable growing states. In fact, the
complex
poles of the S-matrix are related, as it is well known, with unstable physical
states. These poles can then be transformed into
 complex eigenvalues $ z_n$ of the Hamiltonian using standard methods [3]. The
essence of the rigged Hilbert space rather than the Hilbert space,  as the
framework for the quantum states of the system, is clearly exhibited precisely
at this stage: the eigenvalues $z_n$ of a Hermitian operator are  not real
anymore in this extended space \cite{6,11}. If Im $z_n > 0$ then a growing
 prefactor appears in the time evolution of the corresponding eigenvector $ | n
_+\rangle $, giving rise to a growing state belonging to the rigged  Hilbert
space $ \Phi^\times_+ $. On
 the contrary,  if Im $z_n < 0 $ the prefactor is a decaying one, the
corresponding state $\vert n_-\rangle$ is decaying and belongs to another
rigged
Hilbert  space, $
\Phi^\times_-
$. Finally, Im
$z_n = 0
$ corresponds to an ordinary stable state belonging to the ordinary Hilbert
space
$ {\cal H} =
\Phi^{\times}_+ \cap \Phi^{\times}_- $ (more general models contain both,
growing  and decaying states [5,6]).

If $K$ is the Wigner time-reversal operator we have

\begin{equation} K : \Phi^{\times}_- \rightarrow  \Phi^{\times}_+ \ ; K:
\Phi^{\times}_+ \rightarrow \Phi^{\times}_-,
\end{equation}

\noindent since decaying states are transformed into growing states (or
vice-versa) by time-invers\-ion. Then the choice of $\Phi^{\times}_-$
(or $\Phi^{\times}_+$) as our space of quantum states implies that $K$ is not
defined inside
$\Phi^{\times}_-$ (or $\Phi^{\times}_+$), so that irreversibility
naturally appears and therefore the arrow of time also appears in the quantum
regime.

It follows that the choice between $\Phi_-$ or $\Phi_+$ is irrelevant,
since these two objetcs are identical (namely one can be obtained from the
other
by a mathematical symmetry transformation), and therefore the universes, that
we
will obtain with one choice or the other, are also identical and not
distinguishable. Only the names
${\it past}$ and {\it future} or ${\it decaying}$ and ${\it growing}$ will
change but physics is the same and, e.g., we will always have equilibrium
toward the future.

Let us summarize the organization of this paper. Section 2 introduce the
model that we will study. In Section 3 it will be demostrated that the S-matrix
of the model has an infinite set of complex poles (or branch cuts), and how
these
poles are transformed in complex eigenvalues that originate, in turn, damping
factors of unstable decaying states. In Section 4 it is shown how these
damping factors enlarge the set of initial conditions where decoherence occurs.
In Section 5 we will prove that there is correlation in all unstable states.
Finally, we briefly state our conclusions in Section 6. Two Appendixes
complement this work.

%%%%%%%%%%%%%%%%%%%%%%%%%%%%%%%%%%%%%%%%%%%%%%%%%%%%%%%%%%%%%%%%%%%%%%%%%%%%
%%%%%%%%%%%%%%%%%%%%%%%%%%%%%%%%%%%%%%%%%%%%%%%%%%%%%%%%%%%%%%%%%%%%%%%%%%%%
%%%%%%%%%%%%%%%%%%%%%%%%%%%%%%%%%%%%%%%%%%%%%%%%%%%%%%%%%%%%%%%%%%%%%%%%%%%%

\section{The Model}

\setcounter{equation}{0}

Let us consider the model of Sec. 3 of Ref. \cite{10} where a
Robertson-Walker metric is studied (that we will mainly consider in the
flat  case), with a total action $ S = S_g + S_f $, being $ S_g $ the
gravitational action and $ S_f $ the matter action (the usual action of a
spinless massive field $\Phi$). The gravitational action is given by

\begin{equation}S_g = M^2 \int d \eta [ -\frac {1}{2} \dot a^2 -
V(a)],\end{equation}
\noindent

where $M$ is the Planck mass, $\eta$ is the conformal time, $a$ is the
Robertson-Walker scale factor, $\dot a = da/d\eta $, and $ V(a) $ is
the potential function that arises from the spatial curvature, a possible
cosmological constant and eventually a classical matter field. As this last
field is arbitrary, for the sake of simplicity, let us study the case
where the classical matter field is such that $ V(a) = B^2/2 ( 1 - A^2/a^2)$
where $A$ and $B$ are arbitrary constants.

This case is the simplest of all, but we believe that the main features that
we will find will also be presented in more general cases, as we will argue
bellow. The role played by the classical field is completely natural, in the
context of this paper. In fact, we will essentially work using some results of
quantum field theory in curved space-time, where the geometry of space-time is
fixed ``a priori" (namely there is no back-reaction). The classical field is,
precisely, the agency that do this job, fixing a class of possible geometries
(but the properties that we will find will be the same for almost all classical
field).

The Wheeler-DeWitt equation for our Robertson-Walker model is

\begin{equation}\Big[\frac{1}{2M^2}\partial_{a^2}+M^2V(a)-\frac{1}{2}\int_k\bigg(
\partial_{\phi_k^2} -
\Omega_k^2\phi_k^2\bigg)\Big]\Psi(a,\Phi)=0.\end{equation}

Thus after making the WKB ansatz, the Hamilton-Jacobi equation appears as
\cite{10}

\begin{equation}( \frac{dS}{da} )^2 = B^2 ( 1  - \frac{A^2}{a^2}
),\end{equation}

\noindent
where $S$ is the principal Jacobi function. Thus the (semi)
classical time parameter or WKB time $\eta$ is given by

\begin{equation}\frac {d}{d\eta} = \frac{dS}{da} \frac{d}{da}.\end{equation}

Then in our simplified model we have the following class of geometries, in
terms
of this conformal time
$d\eta =a^{-1} dt$ \cite{12}

\begin{equation}a = \pm (A^2 + B^2 \eta^2 )^{\frac{1}{2}} + C \;\end{equation}

\noindent
where $C$ is an arbitrary constant. Using different values for this constant
and different choices for the $\pm$ sign we obtain different
classical solutions (in a more general case many constants would be necessary).
Going now to Ref. \cite{13} (eq.\ (3.113)) we can see that the
semiclassical (or quantum field theory in curved space-time) problem is solved
for all four dimensional, spatially flat, cosmological models with scale factor

\begin{equation}C(\eta) = a^2 = A^2 + B^2 \eta^2 \; \    -\infty < \eta <
\infty
\,
\end{equation}

\noindent
where $ A $ and $ B $ are constants. Then if we consider a massive,
conformally coupled scalar field, the energy function $ \Omega ^2_k $
reads

\begin{equation}\Omega^2_k = m^2 a^2 + k^2 = k^2 + m^2( A^2 + B^2 \eta^2 ) \,
\end{equation}

\noindent
where $m$ is the mass of the quantum matter field and $ k^2 =|\vec k |^2$,
where
$\vec k/a$ is the linear momentum of this field, in the case of flat
space Robertson-Walker universe (or a function of this momentum in the two
other
cases, namely open and  close, being $k$ is a discrete variable in the close
case).  Then (2.7) coincides with the last equation of page 70 of
Ref. \cite{13}.

If we ideally consider the evolution of the universe from $ \eta  \rightarrow
- \infty $ to $ \eta \rightarrow + \infty $ (even if really we would like to
have
only an expanding universe and therefore $\eta \geq 0$, we will discuss
this issue below) and we define the corresponding adiabatic vacua $ |0, in
\rangle $ for $ \eta \rightarrow - \infty $ and $ | 0, out \rangle $ for $ \eta
\rightarrow + \infty $ the Bogolyubov coefficients are (Ref. \cite{13},
eq.\ (3.124))

\begin{equation}\alpha_{kj} = \frac {i (2\pi)^\frac{1}{2} \exp (- \frac{\pi}{4}
   \lambda_k)}{\Gamma[\frac{1}{2}( 1- i \lambda_k) ] } \delta_{kj}
   = \alpha_k \delta_{kj},
   \end{equation}

\begin{equation}
   \beta_{kj} = - i \exp ( - \frac{\pi}{2} \lambda_k ) \delta_{kj}
   = \beta_k \delta_{kj},\end{equation}

\noindent
where $\lambda_k = k^2/Bm + A^2m/B$ and $\delta_{kj} $ is the Kronecker
$\delta$,
for the discrete case, and the Dirac $\delta$, for the continuous one.

Let us comment now on the choice of the vacua since really we would like to
study only the evolution $\eta \geq 0$. The
$|0, {\rm out} \rangle $ vacuum is the adiabatic physical vacuum for $ a
\rightarrow
+ \infty$, where the classical regime must naturally appear, therefore
it is a completely reasonable vacuum. Let us suppose that  the vacuum at $\eta
=
0$ is just $| 0, in \rangle $. This is, of course, a completely
arbitrary choice, that will be discussed in the next section, where we will
introduce a general vacuum at $\eta = 0$ that we will call $| 0, 0 \rangle $.
Anyhow with this arbitrary choice (2.8) and (2.9) are corrects.

%%%%%%%%%%%%%%%%%%%%%%%%%%%%%%%%%%%%%%%%%%%%%%%%%%%%%%%%%%%%%%%%%%%%%%%%%%%%
%%%%%%%%%%%%%%%%%%%%%%%%%%%%%%%%%%%%%%%%%%%%%%%%%%%%%%%%%%%%%%%%%%%%%%%%%%%%
%%%%%%%%%%%%%%%%%%%%%%%%%%%%%%%%%%%%%%%%%%%%%%%%%%%%%%%%%%%%%%%%%%%%%%%%%%%%

\section{The poles of the S-matrix and the unstable quantum states}

\setcounter{equation}{0}

 From (3.46) and (3.47) of Ref. \cite{13}, or more generally from
Sec. 2 of Ref. \cite{14} it can be seen that there is  a pole in the
S-matrix (between  the ``in" and the ``out" Fock spaces) where the
function $\Lambda_{ji} = - i \sum_k \beta_{kj} \alpha^{-1}_{ik}$ has a pole,
namely where $\alpha_{kj} = 0$ ( or $\beta_{kj}$ has a pole). Using (2.8) it
must
be $\alpha_k = 0$ or, which is the same thing, that $ \Gamma[1/2(1
-i\lambda_k)]$ would have a pole. $\Gamma (z)$ has a pole if $z=-n$ (n= 0, 1,
2,..., see, e.g., \cite{15} or \cite{16}) (no poles are produced by the
$\beta$'s
given by (2.9)). Therefore
$S$ has a pole if

\begin{equation}k^2 = m B [-\frac{m A^2}{B}- 2i ( n + \frac{1}{2}
)],\end{equation}

\noindent
and the squared energy, for each pole, reads

\begin{equation}\Omega^2_k = m^2 a^2 + m B [-\frac{m A^2}{B}- 2i ( n +
\frac{1}{2} )].\end{equation}

We  will call this energy $\Omega_k$ simply $\Omega_n$. Thus, we have an
infinite set of unstable states with mean life

\begin{equation}\tau_n = \frac{2^{\frac{1}{2}}\{m^2 (a^2 - A^2) + [m^4 (a^2 -
A^2) + 4 B^2 m^2 (n + \frac{1}{2})^2]^{\frac{1}{2}}\}^{\frac{1}{2}}}{2 B m
(n+\frac{1}{2})}.\end{equation}

Let us observe that the energy and mean life are $a$-dependent. Therefore we
have two possibilities:

1) either we can consider that the in and out states corresponds to $a>>1$,
where
these mean lifes are big but still finite, or

2) we transform all the equations to the non-rescaled case, where the
physical real values are the physical time $t=\int a d\eta$, the physical
energy $\Omega_k/a$ and the physical momentun $\vec k/a$.

We follow the first alternative, and sketch the second one in the Appendix A.

Therefore the universe evolution creates unstable particles as well
as stable ones. Using the standard method explained in Refs. \cite{2} and
\cite{3} we can promote these unstable states to vectors of an adequate rigged
Hilbert space and build a basis of this space with stable modes with real
energies $\Omega_k $ plus unstable modes with complex ``energy" $\Omega_n $
given
by (3.2) (in the open case this procedure is direct, since we have a
continuous spectrum to begin with, in the close case we must use assumption 3
of
Ref. \cite {7}).

This would be the state of affairs if we use the (quite arbitry) vacuum
$|0,in\rangle $ of section 2. In this case we have found an infinite discrete
set of unstable states. What happens if we use a generic (i.e., almost any)
vacuum $|0,0\rangle $ at $ \eta = 0 $? This generic vacuum will be
related to $ | 0,in\rangle $ by some Bogolyubov
coefficients $ \bar{\alpha}_{kj}$, $\bar{\beta}_{kj}$. Then the $
\bar{\bar{\alpha}}$ coefficient relating $ | 0,0\rangle $ to
$ | 0,out\rangle $ reads

\begin{equation}\bar{\bar{\alpha}}_{ik}= \sum_j \bar{\alpha}_{ij} \alpha
_{jk} +\bar{\beta}_{ij}
\beta^*_{jk}= \frac{\bar{\alpha}_{ik} i (2\pi)^{\frac{1}{2}} \exp(
-\frac{\pi}{4}\lambda_k)}{\Gamma [\frac{1}{2}( 1 - i\lambda_k)]}
+ \bar{\beta}_{ik} i \exp( - \frac{\pi}{2} \lambda_k).\end{equation}

The poles are now located where this alpha vanishes. The roots in $k$ of the
corresponding equation, $\bar{\bar{\alpha}}_{ik} = 0 $ can be found only if
we fix the arbitrary coefficients $\bar{\alpha}_{ik}$, $\bar{\beta}_{ik}$.
Of course if these coefficients are fixed in a very particular way the equation
will have no roots. But if the functions $\bar{\alpha}_{ik}$,
$\bar{\beta}_{ik}$ are fixed in a generic (i.e. in almost any) way, the
equation will have a set of complex roots, that correspond to unstable
particles
created by the universe evolution. This statement is equivalent to claim that a
generic S-matrix, for our problem, has infinite numbers of poles or cuts. Let
us
say an infinite set of poles to precise the ideas (cuts will be studied in
Appendix A). Even if we have not, by  now, a rigorous mathematical proof of
this
theorem, we think that we can sketch a reasonable convincing demonstration

Let $ \{ | n, in \rangle \}$, $\{ \vert  0, 0 \rangle \}$, and $ \{\vert m, out
\rangle \}$ be the basis of the Fock spaces corresponding to vacua $| 0,
in\rangle$, $| 0, 0 \rangle$, and $| 0, out\rangle$. The in-out S-matrix, with
an
infinite set of poles, reads

\begin{equation}S_{nm} = \langle n, in | m, out\rangle  = \sum_l \langle n, in|
l, 0 \rangle \langle l, 0 | m, out\rangle\end{equation}

\noindent
where the states $\vert l,0\rangle$ are a complete set.
 From some (infinite) values of $n$ and $m$ we know that $S_{nm}$
has poles. Let us consider one of these values, then the l.h.s. of
(3.4) has also a pole, and theferore one of its terms has a pole. Then,
either:

i) one of the factors inside the summatory of (3.4) has a finite
number of poles and the other one an infinite number of poles, or

ii) both factors have an infinite number of poles.

But (i) must be excluded since, in this case, time evolution $ ( -\infty < \eta
<
0 )$ would be qualitatively different to evolution $( 0 < \eta < +\infty )$ and
this fact would break the time symmetry, which is impossible since evolution
equations, time evolution of $a$, and boundary conditions are time symmetric
with respect $\eta =0$. Then, the $0-out$ matrix $\langle l, 0 | m,
out\rangle$,
corresponding to evolution $\eta \geq 0$, has an infinite number of poles.

We give an alternative demonstration in Appendix B, which is valid for every
evolution and every spatial geometry.

Even if a rigorous proof of these facts would be welcomed we believe that the
reasonings above, and the ones in Appendix B, are quite convincing. Essentially
the periodical nature of
$\bar{\bar{\alpha}}_{ik}$ is inherited from its definition;
$\bar{\bar{\alpha}}_{ik}=(\bar{u}_i , u_k)$ (eq. (3.36), Ref. [13]) where
$\bar{u}_i$ and $u_k$ are two different negative frequency solutions of the
corresponding Klein-Gordon equation. As in flat space-time, these solutions are
functions like $\exp (-i k t)$, they somehow must keep the periodicity in $k$
in curved space-time. The $\alpha$ coeficients always have a periodic
behaviour in the complex plane as it is shown in equations (3.91), (3.124),
(4.60), (4.61), (4.95), (5.41), (5.110), and (5.111) of Ref. [13]. Therefore
the
S-matrix has an infinite and discrete set of complex eigenvalues for almost any
initial condition $\vert 0,0\rangle$.

Now that we know that the 0-out S-matrix has an infinite set of complex poles,
we can find the complex eigenvalues [2,3].

As the \{ $\vert k,out\rangle\}$ basis is complete we have

\begin{equation}\int_k \vert k,out\rangle\langle k,out\vert dk=1,\end{equation}

\noindent
where $\vert k,out\rangle$$ \in {\cal H}$ and the integral means that we must
integrate over the continuous spectrum of energies and other quantum numbers.
Using the standard techniques of Ref. [2,3] we can transform the last equation
in

\begin{equation}\sum_n\vert n,out-\rangle\langle n,out+\vert
+\int_k\vert k,out-\rangle \langle k,out+\vert dk=1,\end{equation}

\noindent
where $\vert n,out-\rangle$, $\vert k,out-\rangle$$\in \Phi_-^\times$ and the
first summatory corresponds to the discrete unstable modes and the integral to
the stable continuous ones.

We choose a $\Phi_-$ test function based in Hardy functions for below, all the
poles will have negative imaginary part and all the unstable states are
decaying
ones, and they will belong to $\Phi^{\times}_-$ (as we already know we can also
make a symmetric choice).

According to Wheeler-DeWitt equation (2.2), the field hamiltonian reads

\begin{equation}h=\frac{1}{2}\int_k\bigg(-\partial_{\phi_k^2}^2+\Omega_k^2
\phi_k^2 \bigg)dk =\int_k\Omega_ka_k^\dagger a_k dk,\end{equation}

\noindent
where $a_k$ and $a_k^\dagger$ are the usual creation and anihilation operators.
 From now, we will always refer to the out case with $a >> 1$ and $h$,
$\Omega_k$, $a_k$, and $a_k^\dagger$
will be $h^{out}$,$\Omega_k^{out}$, $a_k^{out}$, and $a_k^{\dagger out}$. There
are new creation and anihilation operators for the discrete spectrum:
$\bar{a}_n^{out}$, $\bar{a}_n^{\star out}$ and for the continous ones
$\bar{a}_k^{out}$, $\bar{a}_k^{\star out}$ (the definition of $\star$ is given
in
Appendix B). Vectors $\vert n,out-\rangle$ will be created by the repeated
action of $\bar{a}_n^{\star out}$ on $\vert 0,out\rangle$, and vectors
$\vert k,out-\rangle$ will be created by $\bar{a}_k^{\star out}$ analogously.

Therefore $h^{out}$ now reads

\begin{equation}h^{out}=\sum_n\bar{a}^{\star out}_n
\bar{a}^{out}_n+\int_k\Omega_k\bar{a}^{\star out}_k \bar{a}^{out}_k
dk,\end{equation}

\noindent
and we will have

\begin{equation}h^{out}\vert n,out\rangle = \Omega_n n \vert n,out\rangle
,\end{equation}

\noindent
so the evolution of $\vert n,out\rangle$ $\in \Phi_-^\times$ has a damping
prefactor $\exp(-n \eta /\tau_n)$ since $\Omega_n$ has an imaginary commponent.

Thus, as we now have damping factors $\exp(- n \eta /\tau_n)$ in the evolution
equations, it will be very easy to find Lyapunov variables, and in particular a
growing entropy for almost any initial condition as in Refs. \cite{5,6}. This
result must be compared with the one of Ref. \cite{17} where a Luapunov
variable
was found for the universe evolution using the standard methods
\cite{18,19} based in an arbitrary coarse-graining and a particular
(generalized molecular chaos) initial condition. The new result is much more
satisfactory than the old one, since now we have a growing entropy for almost
any initial conditions solving Lochsmidt criticisms.

\section{Decoherence}
\setcounter{equation}{0}

Decoherence is a dissipative process, and we know \cite{20} that it is closely
related to another dissipative phenomenon, that is particle creation from
the gravitational field \cite{21,22,23}. Particle creation has been studied in
the quantum field theory in curved spaces \cite{13,24} as the semiclassical
limit of quantum cosmology. We will restrict ourselves to the semiclassical
approximation to study the decoherence phenomenon.

Decoherence naturally appears in systems where the hamiltonian has complex
eigenvalues, as it is proved in Ref. \cite{25}. Let us consider the formalism
developed in Refs. \cite{9} and \cite{10} to see that, this is also the case,
in the system we are studying. We labelled
the three-geometry with the scalar factor $a$ (the indices $\alpha$ and
$\beta$ simbolizes the choice of the sign and constant in (2.5)), and
$\Phi_N$ is the mode $N$ of the matter field; precisely, we have used $n$ for
the discrete unstable states and
$k$ for the continous stable states; when we will be refering to both kinds of
states, we will call the index
$N$.

The WKB solution of the Wheeler-DeWitt equation reads (\cite{9} eq. (2.8))

\begin{equation}\Psi(a,[\Phi_N])=\exp [iM
S(a)]\chi(a,[\Phi_N]),\end{equation}

\noindent
where $S$ is the principal Jacobi function of (2.3) and
$\chi(a,[\Phi_N])$ can always be written as

\begin{equation}\chi(a,[\Phi_N])=\prod_N \chi_N(\eta ,\Phi_N).\end{equation}

We can obtain $\chi_N(\eta ,\Phi_N)$ by a Gaussian
approximation \cite{10}

\begin{equation}\chi_N(\eta ,\Phi_N) = A_N(\eta) \exp [i \alpha (\eta) -
B_N(\eta)\Phi_N^2].\end{equation}

Functions $A_N(\eta)$ and $\alpha_N(\eta)$ are real while $B_N(\eta)$ is
complex,
precisely $B_N(\eta)=B_{NR}(\eta)+iB_{NI}(\eta)$ and can be obtained solving
the
system
\cite{10}:

\begin{equation}A_N(\eta)=\pi^{-\frac{1}{4}}(2B_{NR}(\eta))^{\frac{1}{2}},
\end{equation}

\begin{equation}\dot{\alpha}_N(\eta)=-B_{NR}(\eta),\end{equation}

\begin{equation}\dot{B}_N(\eta)=-2iB_N^2(\eta)+\frac{i}{2}\Omega_N^2(\eta).\end{equation}

 From [10] or \cite{20} we can learn the conditions for the ocurrence of
decoherence if we use only the real $\Omega_N$, namely the ones of the
continuous spectrum. In Ref. \cite{26} the computation is made only through a
linear approximation of $B(a)$ as a function of $a$. In Ref. [20] decoherence
takes place only in the case where the Bogolyubov coefficients
$\beta_n$ are small and imaginary. In Ref. [10] decoherence takes place unless
the environment is very ordered and fine tuned.  Thus we cannot say that there
is
decoherence for almost any initial condition if the
$\Omega_N$ are all real, like in these works. Let us see, what happens
in our model if we use a basis with infinite complex modes $\Omega_n$ as well
as
real modes
$\Omega_k$.

 From the wave function (4.1), and after the integration on modes of the scalar
field (considered here as the ``environment"), we obtain the following ${\it
reduced}$ density matrix

$$\bar{\rho}_r(a, a') = \exp[-iMS_{\alpha}(a) +iMS_{\alpha}(a')]
\bar{\rho}^{\alpha \alpha}(a,a')$$

$$+ \exp[-i M S_{\alpha}(a) + i M S_{\beta}(a')]\bar{\rho}^{\alpha
\beta}(a,a')$$

$$+\exp[-iM S_{\beta}(a) + iM S_{\alpha}(a')]
\bar{\rho}^{\beta \alpha}(a,a') $$

\begin{equation}+ \exp [-i M S_{\beta} (a) + i M S_{\beta}(a')]
\bar{\rho}^{\beta
\beta}(a,a'),\end{equation}

\noindent
where, as we have said, $\alpha$ and $\beta$ symbolize two different classical
solutions, namely two different choices of the sign $\pm$ and the constant $C$
of
(2.5), and

\begin{equation}\bar{\rho}_r^{\alpha \beta} (a,a') = \prod_N
\bar{\rho}_{rN}^{\alpha \beta}(a,a') = \prod_N \int d\Phi_N
{\chi_N^\alpha}^\ast(\eta ,\Phi_N)\chi_N^\beta(\eta',\Phi_N).\end{equation}

 From (3.20) of Ref. [10], it is

\begin{equation}B_N=-\frac{i}{2}\frac{\dot g_N}{g_N},\end{equation}

\noindent
where $g_N$ is the wave function that represents the quantum state of the
universe being also the solution of
the differential equation

\begin{equation}\ddot g_N+\Omega_N^2 g_N=0,\end{equation}

\noindent
$\Omega_N$ can be the complex energy $\Omega_n$ in our treatment. From (2.6),
(2.7), and (3.1) we know that if the initial sate is $\vert 0,in\rangle$, the
complex energies are

\begin{equation}\Omega_n^2=m^2a^2-m[2iB(n+\frac{1}{2})+mA^2].\end{equation}

In the more general case we use an arbitrary initial state $\vert 0,0\rangle$,
instead of $\vert 0,in\rangle$. From the discussion of Sec. 3, we know that,
in a generic case, an infinite set of complex poles does exist. Then we must
change (3.1) by $k^2=k_n^2$ ($n=0, 1, 2, .....$) where this are the points
where the infinite poles are located in complex plane $k^2$; (3.2) now reads

\begin{equation}\Omega_n^2=m^2a^2+k_n^2.\end{equation}

Let us now see that decoherence takes place if there is an infinite set of
complex modes (even in a more general case than the one of the time evolution
fixed by (2.3), the only one we have studied in great detail above, if we use
the theorem of Appendix B).

Let us consider the asympthotic (or adiabatic) expansion of function $g_N$ when
$a\rightarrow +\infty$ in the basis of the out modes. $g_N$ is the wave
function
that represent the state of the universe, corresponding to the arbitrary
initial
state
$\vert 0,0\rangle$, and its expansion reads

\begin{equation}g_N=\frac{P_N}{\sqrt{2\Omega_N}}\exp [-i\int_0^\eta \Omega_N
d\eta]+\frac{Q_N}{\sqrt{2\Omega_N}} \exp [i \int_0^\eta \Omega_N
d\eta],\end{equation}

\noindent
where $P_N$ and $Q_N$ are arbitrary coefficients showing that $\vert
0,0\rangle$ is really arbitrary.

It is obvious that if all the $\Omega_N$ are real, like in the case of the
$\Omega_k$, (4.13) will have an oscilatory nature, as well as its derivative.
This will also be the behaviour of $B_k$ in (4.9). Therefore the limit of $B_k$
when
$\eta \rightarrow +\infty$ will be not well defined even if $B_k$ itself is
bounded.

But if $\Omega_N$ is complex the second term of (4.13) will have a damping
factor and the first a growing one. In fact, the complex extension of eq.
(4.13) (with $N=k$) reads

\begin{equation}g_n=\frac{P_n}{\sqrt{2\Omega_n}}\exp [-i\int_0^\eta \Omega_n
d\eta]+\frac{Q_n}{\sqrt{2\Omega_n}} \exp [i \int_0^\eta \Omega_n
d\eta],\end{equation}

Therefore when $\eta \rightarrow +\infty$ we have

\begin{equation}B_n
\approx -\frac{i}{2}\frac{\dot{g}_N}{g_N}=\frac{1}{2}\Omega_N.
\end{equation}

Then we have two cases:

i) $\Omega_N=\Omega_k$ $\in {\cal R}^+$ for the real factors. Then we see that
when
$\eta \rightarrow +\infty$, the r.h.s. of (4.8) is an oscillatory function
with no limit in general. We only have a good limit for some particular
initial conditions (as $Q_N=0$ or $P_N=0$  \cite{9,10,20,26}).

ii) $\Omega_N=\Omega_n=E_n-\frac{i}{2}\tau_n^{-1}$ $\in {\cal C}$ for the
complex factors. If we choose the lower Hardy class space $\Phi_-$ to define
our
rigged Hilbert space we will have a positive imaginary part, and there will be
a
growing factor in the first term of (4.13) and a damping factor in the second
one. In this case, for $a\rightarrow +\infty$, we have a definite limit:
$B_n=1/2$ $\Omega_n$.

So we can say nothing about the limit of the real factors (and therefore
nothing in general for the product of these real factors) while the complex
factors have definite limits for every initial conditions, namely for every
$\vert 0,0\rangle$ state.

Therefore let us compute the $\bar{\rho}_{rn}^{\alpha\beta}$ for the complex
factor, since these are the only quantities whose limits we know for sure. So
let
us compute these matrix elements using eq. (2.29) of Ref. [9] or (2.24) of
[10],
namely

\begin{equation}\bar{\rho}_{rn}^{\alpha\beta}(a,a')=\bigg(\frac{4B_{nR}
(\eta,\alpha)B_{nR}(\eta',\beta)}{[B^\ast_n(\eta,\alpha)+B_n(\eta',\beta)]^2}\bigg)^
\frac{1}{4}\exp
[-i\alpha_n(\eta,\alpha)+i\alpha_n(\eta',\beta)],\end{equation}

\noindent
where $\alpha$ and $\beta$ mean that the $B$
refers to these classical solutions. $\bar{\rho}_r^{\alpha\beta}(a,a')$ can be
obtained using (4.8). Let us first compute

\begin{equation}log \vert \bar{\rho}_r^{\alpha\beta}(a,a')\vert =\sum_n
log \vert \bar{\rho}_{rn}^{\alpha\beta}(a,a')\vert .\end{equation}

Now it can be proved that if ${\cal I}m B_n \approx {\cal I}m \frac{1}{2}
\Omega_n \not= 0$,

\begin{equation}\vert \bar{\rho}_{rn}^{\alpha\beta}(a,a')\vert <
1.\end{equation}

In fact, calling $B_n^\ast (\eta ,\alpha)= z = x+iy$ and $B_n(\eta',\beta') =
\zeta = \xi + i y$, we can compute

\begin{equation}\Big\vert \frac{ 4 x \xi}{(x+\xi )^2 + (y+\eta)]^2}\Big\vert
^\frac{1}{4} < \Big\vert \frac{4 x \xi}{\vert x+\xi\vert ^2}\Big\vert
^\frac{1}{4}
\leq 1, \end{equation}

\noindent
since from $\vert x+\xi\vert ^2 \geq 0$ it follows that $4x\xi\leq \vert
x+\xi\vert ^2$. Then all terms of the r.h.s. of (4.17) are negative if
${\cal I}m$ $B_n \not = 0$.

Now from (4.11) or (4.12) we can see that the $B_n(\eta,\alpha)\approx 1/2
\Omega_n(\eta,\alpha)$, corresponding to the discrete complex modes, have all
almost the same asympthotic value when $a \rightarrow +\infty$, therefore all
the
terms of the r.h.s. of (4.17) have almost the same asymthotic value. As they
are all negative and almost equal, the summatory of (4.17) has an asynthotic
value $-\infty$ and therefore $\bar{\rho}_r^{\alpha\beta}(a,a')$ of
(4.8) vanishes only considering the discrete complex modes factors.

Then we have decoherence if $B_n^\ast (\eta,\alpha) \not = B_n(\eta',\beta)$
namely if $\Omega^\ast_n(\eta,\alpha)\not = \Omega_n(\eta',\beta)$, or using
(4.12), if (for an infinite set of $n$) we have

\begin{equation}m^2[\pm (A^2+B^2\eta^2)^\frac{1}{2}+C_\alpha]^2 \not =
m^2[\pm(A^2+B^2{\eta'}^2)^\frac{1}{2}+C_\beta]^2.\end{equation}

So we necessarily have decoherence:

i) for different classical solutions, i.e. for different constants $C_\alpha
\not
= C_\beta$, or different $\pm$ signs, even if the time is the same $\eta =
\eta'$.

ii) for the same classical solutions ($C_\alpha = C_\beta$, and same $\pm$
sign)
if the times $\eta$ and $\eta'$ are different.

Now we can discuss the choice of either $\Phi_-$ or $\Phi_+$ for the test
function spaces. $\Phi_-$ produces the space  $\Phi_-^\times$ of the decaying
states, while $\Phi_+$ will produce the space $\Phi_+^\times$ or growing
states.
Of course one choice becomes the other if we change the $\pm$ sign in (2.5) and
also in all other equations that are a consequence of (2.5). To choose $\Phi_-$
or
$\Phi_+$ corresponds to the choice of the arrow of time in the quantum regime
as
explained in the introducction. Choosing the Gel'fand triplet
$\Phi_-\subset {\cal H}\subset \Phi_-^\times$ is equivalent to say that the
unstable created particles produced by the universe expansion will decay. This
is the motivation of the choice of the arrow of time in the quantum regime.

Nevertheless we can conceive more complex models where growing and decaying
particles could coexist at the same time. Then, we can adopt a more
conservative
attitude. We can consider that the space
$\Phi_+\oplus\Phi_-$ as the space of the test functions with the correponding
space
$\Phi_+^\times
\oplus
\Phi_-^\times$ where there are mixed decaying and growing states. This choice
is
possible in the quantum regime, and therefore there will be no arrow of time in
this regime. But in the classical limit, solution in space
$\Phi_+^\times$ will decohere with the solutions in $\Phi_-^\times$ since they
correspond to different choices of the $\pm$ sign [26].
Therefore for a classical universe either we have a state in $\Phi_+^\times$ or
$\Phi_-^\times$ (which, on the other hand, it is an irrelevant choice). Under
this perspective, we will not have an arrow of time in the quantum regime but
this arrow will appear naturally in the classical regime.

\section{Correlation}
\setcounter{equation}{0}

 From Ref. [10] we can also learn the conditions for the existence of
correlation. But, for the same reason used in Sec. 4, we cannot say that
there is correlation for almost any initial condition. This correlation
depends on the initial conditions, and it can be easily obtained from system
(4.4)-(4.6). In fact for certain initial condition, if the $\Omega_N^2$ are
real, it turns out that
$B_{NR}=0$ and all the conditions for the existence of correlations of Ref.
[10]
 are not fulfilled and therefore there is
no correlation. Precisely, if
$B_{NR}=0$ and
$B_N=iB_{NI}$ and all energies $\Omega_N$ are real (it would be better to
call it $\Omega_k$), (4.6) reads

\begin{equation}\dot{B}_{NI}=2B_{NI}^2+\frac{1}{2} \Omega_N^2,\end{equation}

\noindent
where all the variables are real. Therefore if $B_{NR}=0$ at $a=0$, $B_{NR}=0$
at
every time and there is no correlation. Thus correlation, as decoherence,
depends crucially on the initial condition and there is no correlation for the
above initial condition. Correlation takes place inside each classical solution
and it therefore can be computed using the Wigner function, associated with
$\bar{\rho}_{rn}^{\alpha\alpha}(\eta,\eta')$ \cite{10,27}

\begin{equation}F^{\alpha\alpha}_W(n)(a,P)=\int_{-\infty}^{+\infty}d\Delta
\exp{(-2iP\Delta)} \bar{\rho}_{rn}^{\alpha\alpha}(a-\frac{\Delta}{M},
a+\frac{\Delta}{M}).\end{equation}

\noindent
where $a$, $a'$ $=a\pm \frac{\Delta}{M}$, and $P$ is the canonical momentum.

Nothing new can be said about the real continuous modes, all was already said
in
Ref. [10]. We must only study the complex discrete unstable modes. This is
nevertheless important since, most likely, the universe is in an unstable mode
or more generally in a linear combination of unstables modes (see Ref. [7] and
\cite{28,29}, where the universe is in a ``tunneling" unstable state, i.e. a
typical Gamow vector).

Then we can repeat the reasonings of [10] from (2.24) to (2.28) and, with the
same assumptions we will arrive to this last equation, that now reads

\begin{equation}F^{\alpha\alpha}_W(n)(T,P)\approx
C^2(T)\sqrt{\frac{\pi}{\sigma^2}} \exp\Big[-\frac{(P-M\dot S+\dot
\alpha-\frac{\dot{B}_{nI}}{4B_{nR}})^2}{\sigma^2}\Big].\end{equation}

In the case of the our unstable states we have, for $a(\eta) \rightarrow
+\infty$

\begin{equation}\dot\alpha =-B_{nR}=-\frac{1}{\sqrt
2}\Big[m^2B^2\eta^2+(m^4B^4\eta^4+4m^2B^2(n+\frac{1}{2})^2)^{\frac{1}{
2}}\Big]^{\frac{1}{2}},\end{equation}

\begin{equation}\dot{B}_{nI}=2\sqrt{2}\frac{m^3B^3
\eta}{\Big[m^2B^2\eta^2+\bigg(m^4B^4\eta^4+4m^2B^2(n+\frac{1}{2})^2\bigg)^
{\frac{1}{2}}\Big]^2+4m^2B^2(n+\frac{1}{2})^2},\end{equation}

\noindent
and the inverse of the correlation width of the reduced density matrix is

\begin{equation}\sigma^2=\frac{1}{2}\frac{m^4B^4\eta^2\Big[m^2B^2\eta^2+
(m^4B^4\eta^4+4m^2B^2(n+\frac{
1}{2}))^\frac{1}{2}\Big]^{\frac{1}{2}}}{\Big[m^2B^2\eta^2+
(m^4B^4\eta^4+4m^2B^2(n+\frac{
1}{2}))^\frac{1}{2}\Big]^2+4m^2B^2(n+\frac{1}{2})},\end{equation}

\noindent
if $\eta$ is big: $\vert \dot{B}_R\vert > \vert
\dot{B}_I\vert$.

We can see that when $ \eta\rightarrow +\infty$, $\sigma^2
\rightarrow 0$, and there is a good correlation and the Wigner function is a
Gaussian function, of width $\sigma$, peaked about

\begin{equation}P=M\dot S-\dot
\alpha+\frac{\dot{B}_{nI}}{4B_{nR}},\end{equation}

\noindent
where the first term of the r.h.s. gives the classical result and the last two
are the quantum correlation to the classical trajectory.

On the other hand, in this state we can predict strong correlations
between coordinates and momenta [10], because

\begin{equation}\bigg(M\dot S-\dot\alpha +\frac{B_{nI}}{4B_{nR}}\bigg)^2>>
\sigma^2,\end{equation}

\noindent
(in the preceeding equation we can see that, in the limit of large $\eta$, the
l.h.s. is proportional to $\eta^2$, while $\sigma^2 \sim \eta^{-1}$).

For the generic initial
condition $\vert 0,0\rangle$ we can use (4.12) and we will reach to the same
conclusion. Therefore there is a perfect correlation for the unstable states of
our model. Decoherence and correlation will produce the outcome of an
asymthotic
classical regime in the far future.

\section{Conclusions}

We have demonstrated that, the S-matrix of almost any quantum field theory in
curved spaces model, has an infinite set of poles (or cuts). The presence of
this singularities produces the appearance of unstable states (with complex
eigenvalues) in the universe evolution. The corresponding eigenvectors are
Gamow vectors and produce exponentially
decaying terms as in the Friedrichs model of resonances. But the best
feature of these decaying terms is that they simplify and clarify
calculations.

E. Calzetta and F. Mazzitelli [20] have demonstrated that, under suitable
conditions, the expansion of the universe leads to decoherence if this
expansion
produces particle creation as well. Our unstable states
enlarge the set of initial conditions where decoherence occurs. In fact, the
damping factors (related to the imaginary part of S-matrix's poles), allow that
the interference elements of the reduced density matrix, dissapear for almost
any initial conditions.

Following the reasonings of Ref. [10], we also demostrate that the unstable
states satisfy the correlation conditions, which, with the decoherence
phenomenon, are the origin of the semiclassical Einstein equations.

For simplicity, we assume (as usual) that the state of the
environment can be described by a Gaussian wave function (eq. (4.3)). This is
indeed a restricted class of states [10], but general states could also be
implemented in our formalism. The arbitrary election of the
coefficients $P_N$ and
$Q_N$, in eq. (4.13), shows that the set of initial conditions is really
arbitrary.

Finally, we can say that the existence of unstable states in the universe
evolution (coming from singularities in the Riemann second sheet of the
analytical extension of the S-matrix) can help us to understand the quantum to
classical transition and other dissipative aspects of the universe
evolution.
\vskip 1cm
\section*{Acknowledgments}

We would like to thank I. Prigogine, L. Bombelli, E. Gunzig and
F.D. Mazzitelli for discussions. This work was partially supported by
the Directorate-General for Science, Research and Development of the Commission
of the European Communities under contract ECRU002 (DG-III), by the Institute
Internationaux de Physique et de Chimie Solvay, and the University of Buenos
Aires.

\vskip 1cm
\appendix

\section*{Appendix A}
\renewcommand{\theequation}{A.\arabic{equation}}
\setcounter{equation}{0}

Equation (2.7) reads:

\begin{equation}\Omega_k^2=k^2+m^2a^2,\end{equation}

\noindent
so the relation between the physical energy and momentum is

\begin{equation}\bigg
(\frac{\Omega_k}{a}\bigg )^2=m^2+\bigg (\frac{k}{a}\bigg )^2.\end{equation}

 From (3.1) we know that we will have a resonance if

\begin{equation}\bigg (\frac{\Omega_n}{a}\bigg )^2=m^2+\frac{mB}{a^2}\Big
[-\frac{mA^2}{B}-2i(n+\frac{1}{2})\Big].\end{equation}

Since $a\rightarrow +\infty$ the only way to have resonance for energies
different from $m$, is that at the same time $n \rightarrow +\infty$, so let us
define

\begin{equation}N(n,a)=\frac{n}{a^2},\end{equation}

\noindent
when both $a\rightarrow +\infty$ and $n \rightarrow +\infty$. If $n$ grows in
the proper way, $N(n,a)$ remains finite and we have

\begin{equation}\bigg (\frac{\Omega_k}{a}\bigg )^2=m^2-2imBN.\end{equation}

Then

\begin{equation}\frac{\Omega_k}{a}=\frac{1}{\sqrt{2}}\Big
[m^2+\sqrt{m^4+4m^2B^2N^2}\Big
]^\frac{1}{2}-\frac{i}{\sqrt{2}}\frac{2mBN}{\Big
[m^2+\sqrt{m^4+4m^2B^2N^2}\Big ]^{\frac{1}{2}}}.\end{equation}

So the physical decaying time is

\begin{equation}\tau =\frac{\sqrt{2}a^2}{2mBn}\Big
[m^2+\sqrt{m^4+4m^2B\frac{n^2}{a^4}}\Big ],\end{equation}

where $a>>1$ and therefore also $n >>1$ to get a finite result.

Let us now compute the difference between the square of two subsequent
physical energies

\begin{equation}\bigg(\frac{\Omega_{n+1}}{a}\bigg)^2-\bigg(\frac{\Omega_n}{a}\bigg)^2=-i
\frac{2mB}{a^2}.\end{equation}

When $a\rightarrow +\infty$ this difference vanishes and therefore, most
likely,
we have a cut singularity located at the physical energies

\begin{equation}\frac{\Omega_k}{a}=(m^2-2imBN)^{\frac{1}{2}}.\end{equation}

It is easier to deal with a set of infinite poles that to work with cuts.
Therefore we postpone, further discussion of cuts for future papers.

\section*{Appendix B}
\renewcommand{\theequation}{B.\arabic{equation}}
\setcounter{equation}{0}

Let us demostrate that there are an infinite set of poles or cuts in the
general
case.

 Ordinary anihilation and creation operators are related by the usual
canonical commutation relations

\begin{equation} [ a_{\omega}, a^{\dagger}_{\omega'} ] = \delta (\omega -
\omega').\end{equation}

If we promote the real variable $\omega$ to a complex variable $z$, it is
demonstrated in Ref. \cite{5} that these relations must be subtituted by

\begin{equation}[ a_z , a^\star_{z'} ] = \delta (z - z'),\end{equation}

\noindent where $z$, $z' \in \Gamma$, $\Gamma$ is a contour in the complex
plane
(for details see Ref. \cite{5}) and

\begin{equation}a^\star_z = K a^\dagger_z K^\dagger,\end{equation}

\noindent where $K$ is the Wigner or time-inversion operator. In Refs. \cite{5}
and
\cite{6} it is proved that this operator changes $z$ into $z^\ast$ so really
$a^\star_z$ can be considered as a function of $z^\ast$ rather than $z$.

Now if we make a Bogolyubov transformation among operators labelled by $z$,
the
cannonical conmutation relations (3.6) must be kept invariant. Therefore the
usual relations among Bogolyubov coefficients, such that $\alpha_{ij} =
\alpha_i \delta_{ij}$, $\beta_{ij} = \beta_i\delta_{ij} $, namely

\begin{equation}|\alpha_{\omega}|^2 - |\beta_{\omega}|^2 = 1,\end{equation}

\noindent in the complex case it reads

\begin{equation}\alpha_z \alpha^{\ast}_{z^{\ast}} - \beta_z
\beta^{\ast}_{z^{\ast}} = 1.\end{equation}

Now let us study the functions $\Lambda_z = \beta_z/\alpha_z$. The poles of
these functions originate the poles of the S-matrix. Let us compute:

$$\frac{\vert \beta_z \beta^{\ast}_{z^{\ast}}\vert }{\vert \alpha_z
\alpha^{\ast}_{z^{\ast}}\vert } = \frac{\vert \beta_z
\beta^{\ast}_{z^{\ast}}\vert }{\vert 1+
\beta_z \beta^{\ast}_{z^{\ast}} \vert } = \frac{\vert \zeta\vert }{\vert 1+
\zeta
\vert }$$

\begin{equation} = \frac{\vert \xi + i\eta \vert }{\vert 1 + \xi + i\eta\vert }
=
\frac{(\xi^2  + \eta^2)^{\frac{1}{2}}}{[(1+\xi)^2 +\eta^2]^{\frac{1}{2}}} <
1,\end{equation}

\noindent where $\beta_z \beta^{\ast}_{z^{\ast}} = \zeta = \xi + i\eta $. Then
function
$\beta_z \beta^{\ast}_{z^{\ast}}/\alpha_z\alpha^{\ast}_{z^{\ast}}$ has a
bounded
modulus and therefore it cannot be analytic in all the complex plane if it is
not a constant. But a constant is not a generic function. Therefore function
$\Lambda_z\Lambda^{\ast}_{z^{\ast}}$ must have some singularities some poles
and/or some cuts. If it is a
cut, either $\Lambda_z$ or $ \Lambda_z^\ast$ has a cut and therefore also the
S-matrix, but a cut can be considered (in all the demonstrations of this
paper) like an infinite set of poles and the process is finished and the
theorem
is demostrated. On the other hand let us suppose that the function
$\Lambda_z\Lambda_{z^\ast}^\ast$ has just a pole at $z_0$. Therefore either
$\Lambda_z$ has a pole at $z_0$ or $\Lambda^{\ast}_{z^{\ast}}$ has a pole at
this point. Thus either $S$ has a pole at $z_0$ or at $z^{\ast}_0$, thus in any
case $S$ has at least one pole.

Now as function $\Lambda_z$ has a pole let us say at $z_0$, we can define a
function

\begin{equation}\Lambda^{(1)}_z = \Lambda_z -
\frac{|Res \Lambda_z|_{z_0}}{z-z_0}.\end{equation}

Let us suppose that this function has no poles. We can compute the modulus
$\vert \Lambda^{(1)}_z \Lambda^{(1)\ast}_{z^{\ast}}|$  which obviously is
bounded
since

\begin{equation}|\Lambda^{(1)}_z \Lambda^{(1)\ast}_{z^{\ast}} | <
1,\end{equation}

\noindent if $z$ and $z^{\ast}$ are far enough of $z_0$, and also near to $z_0$
since
$\Lambda^{(1)}_z$ has no pole there, because the only pole of $\Lambda_z$ has
at
$z_0$ and it has been eliminated via the second term of the r.h.s. of
(B.7). Then function $\Lambda_z^{(1)} \Lambda^{(1)\ast}_{z^{\ast}}$ is
analytic and bounded in all the complex  plane and therefore it is a constant.
But again a constant is not a generic function so
$\Lambda_z^{(1)}\Lambda_{z^\ast}^{(1)\ast}$ have some singularities, either a
cut, in which case the proof is finished, or it has a pole at
$z_0^{(1)}$. Now we can define a new function

\begin{equation}\Lambda^{(2)}_z = \Lambda^{(1)}_z - \frac{|Res
\Lambda^{(1)}_z|_{z_0^{(1)}}}{z-z^{(1)}_0},\end{equation}

\noindent and so forth and then prove that $\Lambda_z$, and therefore $S$, have
both an infinite set of poles, or a cut, which can also be considered as an
infinite  continuous set of point-like singularities, for all the issues
considered in this paper. We can also introduce another way to forsee the
presence of this infinite set of poles.

According to what we have just said the equation $\bar{\bar{\alpha}}_{ik}=0$ of
(3.4) has at least a root (since we cannot say that $\bar{\bar{\beta}}_{ik}$
would have any pole in a generic case). But this equation has a ``periodical"
nature (in the complex plane) as we can see if we try to solve a similar
equation like

\begin{equation}\frac{1}{\Gamma(1-z)}= \xi.\end{equation}

Using (1.2.2) of Ref. [16] this equation can be written as

\begin{equation}\frac{1}{\pi}\Gamma(z) \sin \pi z = \xi.\end{equation}

Let us only consider the case $\xi =y$$\in {\cal R}$, $z=x$$\in {\cal R}$,
$x>0$ then

\begin{equation}\frac{1}{\pi}\Gamma(x)\sin \pi x = y.\end{equation}

The function $\Gamma(x)$ is always $>0$ if $x>0$ and it is a growing function.
Then the l.h.s. of (B.12) is a periodical function modulated by a growing
one, then we can see the periodical nature of the problem. Then the function
(B.12) intersects at an infinite set of points the straight line $y=const$.
Therefore (B.12) has an infinite number of roots. As
$\bar{\bar{\alpha}}_{ik}=0$ is a generalization of (B.12) and it has at
least a roots, we can conclude that it has an infinite number of root owing its
periodical nature.


\begin{thebibliography}{99}

\bibitem{1} G.A. Gamow, Z. Phys. ${\bf 51}$, 204 and ${\bf 52}$, 510 (1928)

\bibitem{2} E. C. G. Sudarshan, C. B. Chiu, and V. Gorini, Phys. Rev. D ${\bf
18}$, 2914 (1978)

\bibitem{3} A. Bohm, {\it Quantum Mechanics: Foundations and Applications}
(Springer-Verlag, Berlin) (1986)

\bibitem{4} I. Antoniou and I. Prigogine, {\it Dynamics and Intrinsic
Irreversibility}, Proc. Int. Symp. on Conceptual Tools for Understanding
Nature,
Trieste (1990)


\bibitem{5}  M. Castagnino, F. Gaioli,  and E. Gunzig, {\it Cosmological
Features of Time Asymmetry}, preprint IAFE to be submitted to Foundations of
Cosmic Physics (1994)


\bibitem{6} M. Castagnino and  R. Laura, {\it The Cosmological Essence of Time
Asymmetry}, Proceeding SILARG VIII, ed. by W. Rodrigues (World Scientific,
Singapure) (1994)

\bibitem{7} M. Castagnino, E. Gunzig, P. Nardone, I. Prigogine, and  S. Tasaki,
{\it Quantum Cosmology and Large Poincar\'e Systems}, Fundamental Papers on
Theoretical Physics, ed. M. Namiki, AIP, 1994.


\bibitem{8} M. Castagnino, F. Gaioli, and D. Sforza, {\it Irreversible
Evolutions in Quantum Cosmology}, Anales AFA, vol. 5 (1993)

\bibitem{9} A. Gangui, F. D. Mazzitelli, and M. Castagnino, Phys. Rev. D ${\bf
43}$, 1853 (1991)

\bibitem{10} J. P. Paz and S. Sinha, Phys. Rev. D ${\bf 44}$, 1089 (1991)

\bibitem{11} T. Petrosky, I. Prigogine, and S. Tasaki, Physica A ${\bf 173}$,
175
(1991)

\bibitem{12} J. Audretsch and G. Sch$\ddot a$fer, Phys. Lett., ${\bf 66A}$, 459
(1978)

\bibitem{13} B. Birrell and  P. C. W. Davies, {\it Quantum Field Theory in
Curved Space}, (Cambridge Univ. Press, Cambridge, England) (1982)

\bibitem{14} N. B. Birrell and J. G. Taylor, J. Math. Phys. ${\bf 21}$, 1740
(1980)

\bibitem{15} H. Hochstadt, {\it The Functions of Mathematical Physics }, (Dover
Pub. Inc., New York) (1972)

\bibitem{16} N. N. Lebedev {\it Special Functions and their Application},
(Dover
Publ. Inc., New York) (1971)

\bibitem{17}  E. Calzetta, M. Castagnino, and R. Scoccimarro, Phys. Rev. D
${\bf
45}$, 2806 (1992)

\bibitem{18} R. W. Zwanzig, {\it Statistical Mechanics of Irreversibility}, in
Quantum Statistical Mechanics; ed. P. Meijer (Gordon and Breach, New
York)(1966)

\bibitem{19} R. W. Zwanzig, J. Chem. Phys. {\bf 33}, 1338 (1960)

\bibitem{20} E. Calzetta and F. Mazzitelli, Phys. Rev. D ${\bf 42}$, 4066
(1990)

\bibitem{21}B. L. Hu and E. Calzetta, Phys. Rev. D${\bf 40}$, 656 (1989)

\bibitem{22}B. L. Hu, Physica A ${\bf 158}$, 399 (1989)

\bibitem{23}L. Parker, Phys. Rev. Lett. ${\bf 21}$, 562 (1968); Phys. Rev.
${\bf 183}$, 1057 (1969)

\bibitem{24}L. Parker, in {\it Asymptotic Structure of Spacetime}, eds. F.
Esposito and L. Witten, (New York, Plenum) (1977)

\bibitem{25} M. Castagnino, F. Gaioli, and F. Lombardo, {\it Rigged Hilbert
Space Approach to Brownian Motion Problem}, preprint IAFE (unpublished)

\bibitem{26} J. Halliwell, Phys. Rev. D ${\bf 39}$, 2912 (1989)

\bibitem{27} J. Halliwell, Phys. Rev. D ${\bf 34}$, 3626 (1987)

\bibitem{28} A. Vilenkin, Phys. Rev.  D ${\bf 33}$, 3560 (1986)

\bibitem{29} A. Vilenkin, Phys. Rev.  D ${\bf 37}$,  888 (1988)


\end{thebibliography}
\end{document}